\documentclass[aps,prb,floatfix,twocolumn,showpacs]{revtex4}
\usepackage{epsfig, amsmath, amssymb}

\begin{document}

% Patch REVTeX to prevent BibTeX from seeing endnotes as citations
% Insert just after REVTeX is loaded
%\makeatletter
%\let\@ORGREVTEXendnotemark\@endnotemark
%\let\@ORGREVTEX@makefnmark@cite\@makefnmark@cite
%\def\@endnotemark{\bgroup\@fileswfalse\@ORGREVTEXendnotemark\egroup}
%\def\@makefnmark@cite{\bgroup\@fileswfalse\@ORGREVTEX@makefnmark@cite\egroup}
%\makeatother

%\twocolumn[
\hsize\textwidth\columnwidth\hsize\csname@twocolumnfalse\endcsname

\title{Spin-dependent tunneling into an empty lateral quantum dot}

\author{Peter Stano$^{1,2}$ and Philippe Jacquod$^1$}
\affiliation{$^1$Physics Department, University of Arizona, 1118 E 4th Street, Tucson, Arizona 85721, USA\\
$^2$Institute of Physics, Slovak Academy of Sciences, Bratislava 845 11, Slovakia}

\vskip1.5truecm
\begin{abstract}
Motivated by the recent experiments of Amasha {\it et al.}
[Phys. Rev. B {\bf 78}, 041306(R) (2008)],
we investigate single electron tunneling into an empty quantum dot in presence
of a magnetic field. We numerically calculate the tunneling rate from a
laterally confined, few-channel external lead into the lowest orbital state
of a spin-orbit coupled quantum dot. We find two mechanisms leading to 
a spin-dependent tunneling rate. The first originates from different 
electronic $g$-factors in the lead and in the dot, and favors the tunneling into the
spin ground (excited) state when the $g$-factor magnitude is larger (smaller) in
the lead. The second is triggered by spin-orbit interactions via the
opening of off-diagonal spin-tunneling channels. It systematically favors
the spin excited state. For physical parameters corresponding to
lateral GaAs/AlGaAs heterostructures and the experimentally
reported tunneling rates, both mechanisms lead to a discrepancy of 
$\sim$10\% in the spin up vs spin down tunneling rates. 
We conjecture that the significantly larger discrepancy 
observed experimentally originates from the enhancement of the $g$-factor 
in laterally confined lead. 
\end{abstract}
\maketitle

\section{Introduction}

Spintronics uses the spin, rather than the charge degree of freedom
of electrons for information processing.\cite{fert2008:RMP,grunberg2008:RMP} 
This requires the ability to 
create, manipulate, and detect spin currents and accumulations, all tasks
for which semiconductors seem especially promising.\cite{fabian2007:APS} 
Experiments on few-electron quantum dots in a Zeeman field
have shown how to manipulate two-electron spin states by
electrically controlling the exchange interaction in double-dot systems,
\cite{hanson2007:RMP,taylor2007:PRB} 
and how to convert electronic spin orientations into 
electrostatic voltages.\cite{elzerman2004:N, pfund2009:PRB, hanson2005:PRL}
More recently, Amasha {\it et al.}\cite{amasha2008:PRB} reported 
a strong and unexpected 
spin dependence of the tunneling rate into an empty quantum dot 
in a Zeeman field. In these experiments 
a lateral quantum dot is tunnel-coupled to a narrow
external lead. Charge sensing allows to measure the time it takes
for an electron to enter the dot after an electric pulse has brought 
the lowest orbital level below the Fermi energy in the lead. Once this
level is Zeeman-split the tunneling rates into the ground state and into the
first spin-excited level can be extracted. Changing the magnetic field and the
geometry of the dot, the ratio $\chi$ of the tunneling 
rates for the two spin states was found to vary between $\chi=1$
(symmetric tunneling) to $\chi \ll 1$ (negligible tunneling into the 
spin excited state). It was, in particular, found that 
$\chi \rightarrow 0$ at large in-plane 
magnetic field. This is rather intriguing as one 
expects tunneling rates to depend mostly on the orbital structure of 
the wavefunction, over which a Zeeman field has no effect. 
Because of the importance
that such an effect might have
for quantum dot spintronics, these experimental results need to be better
understood. This is one of our objectives
in this paper.

In III-V semiconductor heterostructures, the spin-orbit interactions are the usual suspect behind any spin-dependent effect. Several mechanisms for spin injection in tunneling structures with spin-orbit interactions have been considered, most
notably based on
resonant enhancement of weak spin-orbit effects,\cite{voskoboynikov1999:PRB, voskoboynikov2000:JAP, glazov2005:PRB}
spatial modulation of the spin-orbit interactions strength,\cite{voskoboynikov1998:PRB,rozhansky2008:PRB}
spin-orbit induced mass renormalization,\cite{perel2003:PRB} 
lateral confinement,\cite{streda2003:PRL, silvestrov2006:PRB,tkach2009:JPCM}
or electron momentum filtering.\cite{fujita2008:JPCM}
Here, we extend these investigations and study electronic 
tunneling into an empty quantum dot from a single external lead in
presence of an external magnetic field.
We uncover two, so far neglected mechanisms for elastic 
spin-dependent tunneling. 
The first one appears when tunneling occurs between an extended and a confined
region with different $g$-factors. Tunneling being elastic, the splitting of
the orbital states in the confined region determines the energy of the 
tunneling electron in the continuum. Because of the discrepancy in the
$g$-factors, however, the continuum 
wavevectors of the tunneling electrons depend on the spin orientation, thus
the tunneling rate becomes spin-dependent. Tunneling into the ground-state
(first spin-excited state)
is favored if the $g$-factor magnitude is larger (smaller) in the lead.
The second mechanism arises because
spin-orbit interactions are effectively weaker in the low-energy spectrum
of a confined region than in the continuum. The direction of the effective
Zeeman field (the sum of the magnetic and spin-orbit fields)
seen by the electron in the lead and in the dot are thus different, and this
opens spin-non-conserving tunneling channels.
Because this second mechanism 
systematically favors tunneling into the spin-excited state, and 
because our numerical investigations estimate a discrepancy of
$\sim$10\% at most in the tunneling rates for realistic parameters, 
we conclude that it plays only a marginal role in the experiments of
Ref.~\onlinecite{amasha2008:PRB}, where tunneling rates into
the ground-state are systematically larger and discrepancies up to almost 100\% are observed. 
In contrast, we conjecture that such large spin-anisotropies in tunneling arise from
the first mechanism when the $g$-factor in the lead
is substantially larger than the dot and the bulk value. Strongly 
enhanced $g$-factors have been observed in the lowest
conduction subbands of quantum point contacts.\cite{thomas1996:PRL,chen2009:PRB,koop2007:JSNM}
They are usually attributed to the electron-electron Coulomb interactions.\cite{pallecchi2002:PRB} We expect that a similar enhancement occurs in narrow, laterally confined leads.

To model the experiment, we consider a two dimensional dot divided by a barrier from a semi-infinite lead of finite width. The potential along the lead axis is sketched in Fig.\ref{fig:mechanisms}a, with Fig.\ref{fig:mechanisms}b giving
a top view of the model structure. We assume that tunneling is elastic, 
which has been found to be the case experimentally.\cite{amasha2007:PRL} 
It is easy to understand how a spin asymmetry in tunneling
can arise. Consider first that there is no spin-orbit interaction. Applying 
an in-plane magnetic field Zeeman-splits electronic levels. 
Tunneling being elastic, the tunneling rate is spin independent if the
$g$-factor is constant, because the barrier height is the same for both spin species. This is sketched in 
Fig.~\ref{fig:mechanisms}c. If on the other hand 
the Zeeman splitting is larger in the dot, the barrier height for the spin excited 
state becomes effectively smaller, thus tunneling into that state is 
faster than into the lowest spin state. The situation is depicted
in Fig.~\ref{fig:mechanisms}d. 
Consider next that spin-orbit interactions are present. They 
are effectively weaker in the dot than in the lead\cite{aleiner2001:PRL,levitov2003:PRB} and because of this,
tunneling can be accompanied by spin-flips in presence
of a Zeeman field. The net result is that spin off-diagonal tunneling channels
open up. Below we show that this leads to a systematically larger 
rate into the spin-excited state in the dot. This mechanism is
sketched in Fig.~\ref{fig:mechanisms}e.

The article is organized as follows. In Sec.~II we present our model for the dot and lead states, and for the tunneling rate. We analyze separately the $g$-factor inhomogeneity and the spin-orbit interactions mechanisms in Secs.~III and IV, respectively. We conclude in Sec.~V. In the Appendix we compare the tunneling formula we use with three common alternatives in a simplified model, to demonstrate its generality.

\section{Model}

We compute tunneling rates using the method of Ref.~\onlinecite{gurvitz1987:PRL}, which to leading order gives
\begin{equation}
\Gamma_{\sigma}=\frac{2\pi}{\hbar}\sum_{c, k, \sigma^\prime} |\langle \Phi^{\rm dot}_{\sigma} | \delta \hat{V} | \Psi^{\rm lead}_{c,k,\sigma^\prime} \rangle|^2 \delta(E^{\rm dot}_{\sigma}-E^{\rm lead}_{c, k, \sigma^\prime}).
\label{eq:fermis}
\end{equation}
It describes an elastic transition from the set of lead states labeled by the transverse channel $c$, longitudinal wavevector $k$, and spin $\sigma^\prime$, into the lowest orbital state of the dot with spin $\sigma$. The effective transition potential $\delta \hat{V}$, is defined as the difference of the potential of the isolated dot and a dot with a nearby lead, see Fig.~\ref{fig:mechanisms}a. The tunneling formula, Eq.~(\ref{eq:fermis}), is discussed in the Appendix, where we additionally show that in one dimension it is equivalent to other, standardly used tunneling formulas.

\begin{figure}
\centerline{\psfig{file=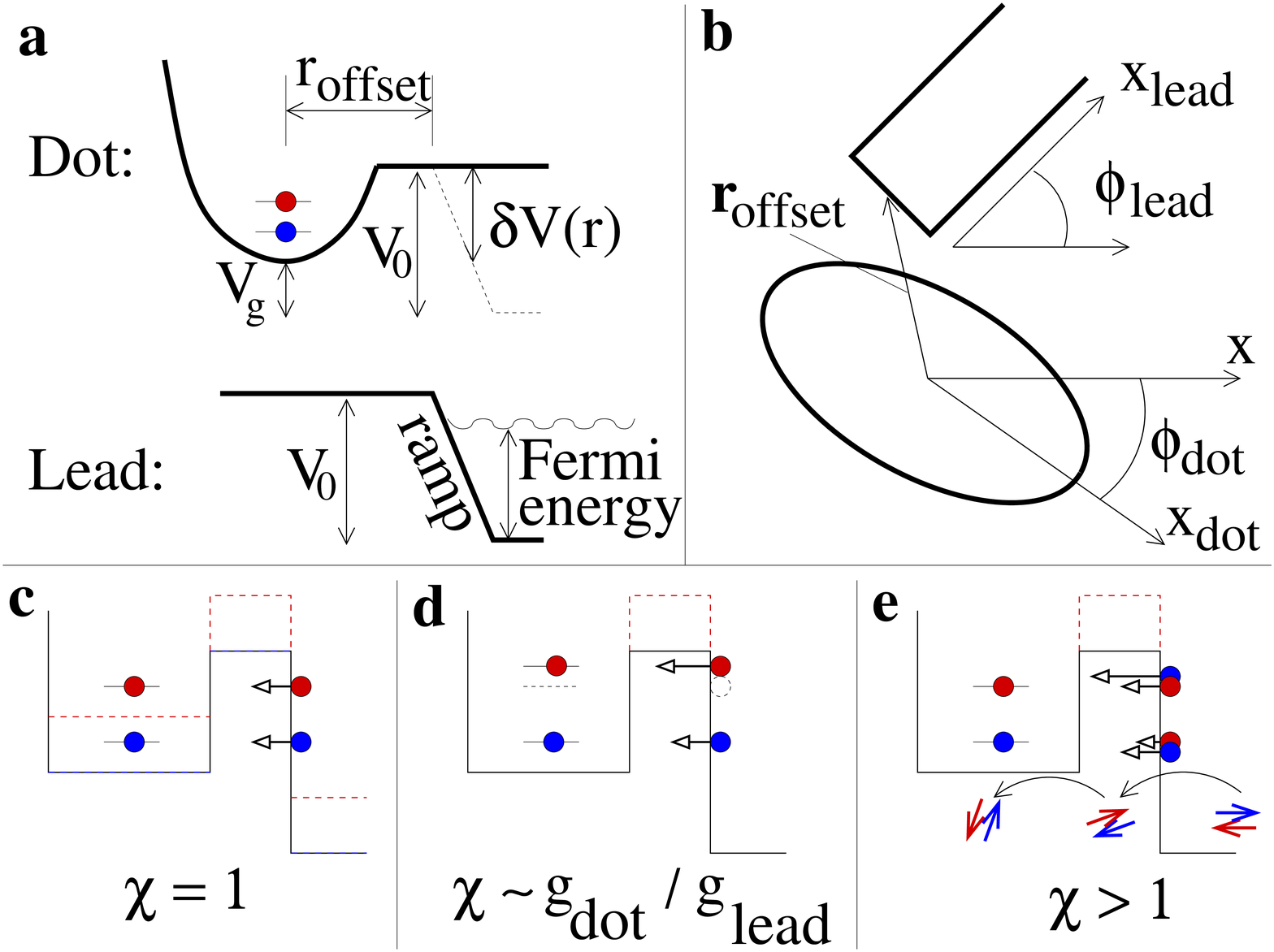,width=1\linewidth}}
\caption{(Color online) (a) Potential profile of the dot and the semi-infinite lead when considered separately. (b) Top view of the dot-lead structure. (c)-(e) Schematics of the tunneling mechanisms. (c) Symmetric tunneling. (d) Asymmetric tunneling: a larger (smaller) $g$-factor in the lead enhances the tunneling into the ground (excited) state. (e) Asymmetric tunneling:
spin-orbit interactions open the off-diagonal spin tunneling channels, which always favors tunneling into the excited state.}
\label{fig:mechanisms}
\end{figure}

The lead and the dot wavefunctions in Eq.~(\ref{eq:fermis}) are those of the two subsystems considered separately. We take them as eigenstates of the following Hamiltonian:
\begin{equation}
H=\frac{\hbar^2}{2m}{\bf \hat{k}}^2 + \frac{g}{2}\mu_B {\bf B}\cdot \boldsymbol{\sigma} + H_{\rm so} + V({\bf r}),
\label{eq:Hamiltonian}
\end{equation}
with appropriate boundary conditions. These are that the dot wavefunctions are zero at infinity, while the extended states of the lead are fixed by the energy, the transverse channel, and the spin of the incoming wave. In Eq.~(\ref{eq:Hamiltonian}), $m$ is the electron effective mass, ${\bf \hat{k}}=-{\rm i}\boldsymbol{\nabla}$, ${\bf B}$ is 
an in-plane magnetic field forming angle $\phi_{\bf B}$ with the crystallographic $x$ axis. It couples to electronic spins
via the Land\'e $g$-factor and the Bohr magneton $\mu_B$, and  
$\boldsymbol{\sigma}=(\sigma_x, \sigma_y, \sigma_z)$ is 
a vector of Pauli matrices. The spin-orbit interactions Hamiltonian $H_{\rm so}$ includes the Bychkov-Rashba, linear Dresselhaus, and cubic Dresselhaus terms\cite{fabian2007:APS}
\begin{subequations}
\begin{eqnarray}
H_{\rm br} &=& \frac{\hbar^2}{2 m l_{\rm br}} (\hat{k}_y \sigma_x - \hat{k}_x \sigma_y),\\
H_{\rm d} &=& \frac{\hbar^2}{2 m l_{\rm d}} (-\hat{k}_x \sigma_x + \hat{k}_y \sigma_y),\\
H_{\rm d3} &=& \gamma_c (\hat{k}_x \hat{k}_y^2 \sigma_x - \hat{k}_y \hat{k}_x^2 \sigma_y),
\end{eqnarray}
\label{eq:spin-orbit}
\end{subequations}
which we write as a momentum-dependent magnetic field $H_{\rm so}=(g/2)\mu_B {\bf B_{\rm so}}({\bf \hat{k}})\cdot \boldsymbol{\sigma}$. 
Here, we adopt the two-dimensional approximation, where only the lowest subband of the perpendicular confinement is occupied. In this case, the in-plane magnetic field has no orbital effect as long as the magnetic length, $\sqrt{2 \hbar/e B}$ is larger than the perpendicular extension of the two-dimensional electron gas (2DEG). This conditions is fulfilled here, since even at the largest magnetic fields of 7.5 Tesla we consider, the magnetic length is $\sim 13$ nm. Moreover, the strengths of the two Dresselhaus interactions are related by $\hbar^2/2 m l_{\rm d} = \gamma_{\rm c} \langle k_z^2 \rangle$, with $\langle k_z^2 \rangle$ being the quantum mechanical variance of $k_z$ in the lowest subband in the perpendicular confinement of the two-dimensional electron gas.

Electronic confinement is described by the potential $V({\bf r})$.
The semi-infinite lead is laterally confined by infinite potential barriers
a distance $w_{\rm lead}$ apart.  Close to the dot, it is terminated by a 
potential barrier linearly reaching a magnitude $V_0$ 
over a distance $L$ (we call this the ``ramp''),
\begin{equation}
V^{\rm lead}(x_{\rm lead})=\left\{ 
\begin{tabular}{ll}
$V_0,$&$x_{\rm lead}<0$,\\
$(1-x_{\rm lead}/L)V_0,$&$0<x_{\rm lead}<L$,\\
0,& $L<x_{\rm lead}.$
\end{tabular}
\right.
\end{equation}
The lead coordinate system is offset by ${\bf r}_{\rm offset}$ with respect to the dot center and rotated by an angle $\phi_{\rm lead}$ with respect to the crystallographic coordinates $x$, and $y$. This is shown in Fig.~\ref{fig:mechanisms}b.

For the dot confinement we take an anisotropic linear harmonic oscillator, bounded from above by $V_0$ in the lead area [see Figs.~\ref{fig:mechanisms}a and \ref{fig:mechanisms}b]
\begin{eqnarray}
V_0^{\rm dot}({\bf r})&=&\frac{\hbar^2}{2 m} \left( x_{\rm dot}^2 l_x^{-4} +  y_{\rm dot}^2 l_y^{-4}\right)+V_g,\\
V^{\rm dot}({\bf r})&=&\left\{ 
\begin{tabular}{ll}
$V_0^{\rm dot}({\bf r})$&$,{\bf r} \notin {\rm lead}$,\\
${\rm min} \{V_0^{\rm dot}({\bf r}), V_0\} $&$,{\bf r} \in {\rm lead}.$
\end{tabular}
\right.
\end{eqnarray}
The main axes of the dot are rotated with respect to the crystallographic axes by an angle $\phi_{\rm dot}$. The two independent confinement lengths $l_x$ and $l_y$ allow us to consider dot deformations as in the experiment of Ref.~\onlinecite{amasha2008:PRB}. The potential $V_g$ is used to offset the dot with respect to the lead. The model has no particular spatial symmetry.

We use parameters typical for GaAs heterostructures. When not varied, their values are 
$m=0.067$, $\gamma_c=27.5$ eV\AA$^3$, $g=-0.39$, $l_d=0.63\mu$m, 
$l_{br}=2.4\mu$m,\cite{footnote3}
and a Fermi energy in the lead $E_{\rm F} = 8$ meV.
The geometry is set to qualitatively 
correspond to the description of the experimental
setup in Refs.~\onlinecite{amasha2008:PRB} and \onlinecite{amasha2008:PRL}, $l_x=l_y=24$ nm,
corresponding to an orbital excitation energy $E_{\rm orb}=2$ meV,
$V_0=12$ meV, $L=240$ nm, $\phi_{\rm lead}=-\pi/4$, $\phi_{\rm dot}=\pi/4$, and $\phi_{\bf B}=3\pi/4$.
We offset the dot by $V_g \approx 6$ meV to align the lead Fermi energy with 
the dot spin excited state. 
We offset the lead with respect to the dot by $y_{\rm offset}=24$ nm and we 
take $w_{\rm lead}=72$ nm, giving two open transverse channels. Finally, we use $x_{\rm offset}$ as a free 
parameter, which we always adjust to keep the total tunneling rate into 
the spin-excited state at 200 Hz. Below we vary all the parameters individually, which allows us to identify which are relevant for spin-dependent tunneling. Because we consider tunneling into an empty dot, we neglect electron-electron interactions in the dot (beyond charging effects which prohibit double occupancy of the dot), but incorporate them in a renormalized $g$-factor in the lead.

We define the asymmetry $\chi$ in spin tunneling as the ratio of the tunneling rates into the two lowest Zeeman split dot states
\begin{equation}
\chi = \Gamma_\downarrow/\Gamma_\uparrow. 
\end{equation}
Symmetric tunneling corresponds to $\chi=1$, while $\chi$ smaller(larger) than one means that tunneling into the ground(excited) spin state is faster.

\section{Spatially dependent $g$-factor}

As a first source of tunneling asymmetry, we consider the inhomogeneity in the $g$-factor and neglect spin-orbit interactions for the time being. A slight difference of the $g$-factor in the dot and the lead is not unexpected, as in semiconductor quantum dots the $g$-factor depends on the magnetic field, state energy, or wavefunction penetration into the barrier along the growth direction.\cite{hanson2003:PRL,falko2005:PRL,kiselev1998:PRB,doty2006:PRL} Reference \onlinecite{amasha2008:PRL} reported mesoscopic fluctuations of the 
$g$-factor in a quantum dot of a relative magnitude $\sim 20\%$, see Fig.~3a there.

\begin{figure}
\centerline{\psfig{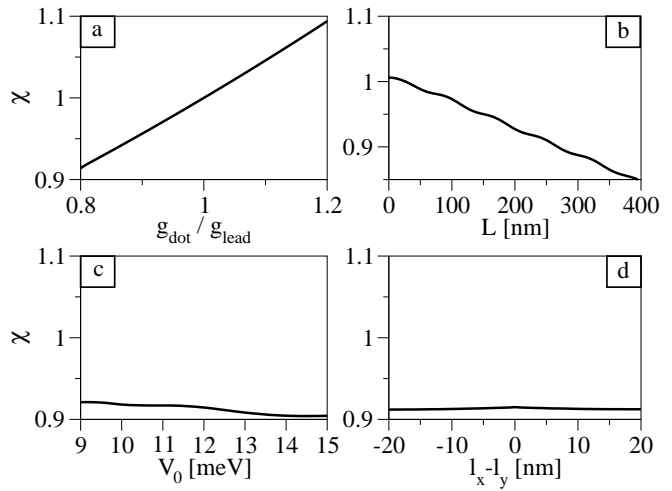}}
\caption{Spin tunneling asymmetry due to the $g$-factor inhomogeneity varying (a) the $g$-factor difference, (b) the length $L$ of the 
barrier ramp, (c) the barrier height $V_0$, and (d) the dot asymmetry ($l_x-l_y$). When not varied, the parameters are $g_{\rm lead}=-0.39$, $g_{\rm dot}=-0.32$, $V_0=$12 meV, $L=240$ nm, and $l_x-l_y=0$. In all panels, the tunneling rate into the ground state has been kept fixed at 200 Hz.}
\label{fig:g-factor}
\end{figure}
We first discuss qualitatively how a $g$-factor difference results in $\chi \neq 1$. According to Eq.~(\ref{eq:fermis}), the tunneling rate from a lead state $\sigma^\prime$ to the dot state $\sigma$ scales with 
\begin{equation}
\Gamma_{\sigma^\prime \to \sigma} \sim |\langle \Phi^{\rm dot}_\sigma | \Psi^{\rm lead}_{\sigma^\prime} \rangle|^2 \sim \exp(-2 k_{\sigma^\prime \to \sigma } L) |\langle \xi^\sigma | \xi^{\sigma^\prime} \rangle|^2 .
\label{eq:WKB}
\end{equation}
The orbital overlap depends on the wavevector
\begin{equation}
k_{\sigma^\prime \to \sigma} = \sqrt{2m [(V_0 + \epsilon_{\sigma^\prime}^{\rm lead}) - (E_0 + \epsilon_{\sigma}^{\rm dot} ) ] / \hbar^2}, 
\label{eq:wavevector}
\end{equation}
where we split the energy levels in the dot 
into the orbital and Zeeman contributions, 
$E_0+\epsilon_{\sigma}^{\rm dot}$, and $\epsilon_{\sigma^\prime}^{\rm lead}$ is
the Zeeman energy in the lead. In the absence of spin-orbit interactions, $\langle \xi^\sigma | \xi^{\sigma^\prime} \rangle=\delta_{\sigma\sigma^\prime}$. 
If the $g$-factor is the same everywhere, there is no tunneling asymmetry
\begin{equation}
\chi \sim \frac{\exp(-2 k_{\downarrow\to\downarrow} L)}{\exp(-2 k_{\uparrow \to \uparrow} L)}=1,
\end{equation}
because $\epsilon^{\rm lead}_\sigma=\epsilon^{\rm dot}_\sigma$. If, however, the Zeeman energies in the dot and the barrier differ, we get
\begin{equation}
\chi \sim \exp\left(-2 L \sqrt{\frac{m}{\hbar^2(V_0-E)}}(g_{\rm lead}-g_{\rm dot})\mu_B B \right).
\label{eq:xi g factor approximation}
\end{equation}
A $g$-factor larger in magnitude in the lead gives a faster tunneling into the ground state, as observed in Ref.~\onlinecite{amasha2008:PRB}. However, assuming a dot $g$-factor reduction of 20\%, we estimate $\chi \sim 0.95$ at 7.5 Tesla, an asymmetry much smaller than observed.

We numerically evaluate $\Gamma_\sigma$ from Eq.~(\ref{eq:fermis}) and plot $\chi$ in Fig.~\ref{fig:g-factor}. Figure \ref{fig:g-factor}a confirms that the tunneling into the ground (excited) state is preferred, if the $g$-factor magnitude is larger (smaller) in the lead than in the dot. We next show in Fig.~\ref{fig:g-factor}b that the asymmetry is larger for tunnel barriers with larger ramps. For a dot-lead $g$-factor difference of 20\%, $\chi \approx 0.9$ for a barrier with $L \approx 200$ nm,
which qualitatively corresponds to the experimental setup of 
Ref.~\onlinecite{amasha2008:PRB}.
There is a slight dependence of $\chi$ on $V_0$ shown in Fig.~\ref{fig:g-factor}c. The trend somehow contradicts Eq.~(\ref{eq:xi g factor approximation}), because  for low barriers, the dot wavefunction leaks out into the lead channel, and this is not captured by the approximations made in Eq.~(\ref{eq:WKB}). Finally, Fig.~\ref{fig:g-factor}d shows that deforming the dot leaves $\chi$ practically unchanged. This is expected, because the symmetry of the ground state depends
only weakly on details of the confinement potential. Most importantly, these numerical results on a two-dimensional model confirm the qualitative picture based on a one-dimensional toy model given above. For variations in the dot's
$g$-factor up to 20\%, the relative tunneling asymmetry does not exceed $\sim10$\% and the symmetry of the confinement potential of the dot plays no role. We conclude that 
the reported fluctuations in the dot's $g$-factor variations can not explain 
the reported large spin-asymmetry in tunneling.

\begin{figure}
\centerline{\psfig{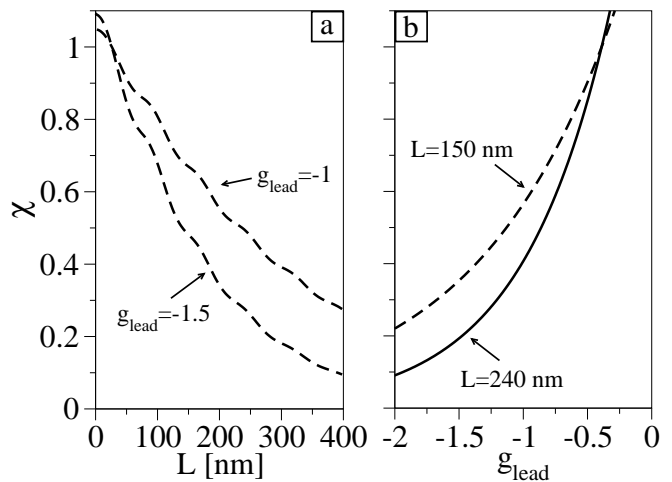}}
\caption{Spin tunneling asymmetry due to the enhancement of the 
$g$-factor in the lead as a function of (a) the barrier length for g-factor values similar to those reported in Refs.~\onlinecite{thomas1996:PRL} and \onlinecite{chen2009:PRB} and (b) the lead g-factor for two fixed barrier lengths.}
\label{fig:g-factor2}
\end{figure}

A small $\chi$ requires a large argument in the exponential of Eq.~(\ref{eq:xi g factor approximation}), which can be reached for a large difference in Zeeman energy or $k L \gg 1$. The barrier width $L$ is limited by the tunneling rate magnitude (200 Hz), and therefore cannot exceed few hundreds of nanometers. We just argued that variations in the dot's $g$-factor cannot account for the observed $\chi$, if $g_{\rm lead}=-0.39$, which is the standardly accepted value for the two dimensional electron gas in
GaAs/AlGaAs heterostructures. We now argue that only an enhancement of the lead $g$-factor can generate a large tunneling asymmetry. A strong enhancement of the $g$-factor, up to $|g|=3$ at magnetic field of 16 Tesla, has been reported in Refs.~\onlinecite{thomas1996:PRL} and \onlinecite{chen2009:PRB} in the lowest conduction subbands of a quantum point contact.
It is reasonable to expect that a similar effect occurs in the narrow leads of Ref.~\onlinecite{amasha2008:PRB}. Two data sets are shown in Fig.~\ref{fig:g-factor2}a for values of the lead $g$-factor as reported in Ref.~\onlinecite{thomas1996:PRL}. The tunneling asymmetry is now strongly enhanced and comparable to the experiment at reasonable barrier lengths. Alternatively, Fig.~\ref{fig:g-factor2}b shows the tunneling asymmetry at two fixed barrier lengths. The $g$-factor enhancement required for a substantial asymmetry is still reasonable, given that enhancement factors of up to 50 were reported in Ref.~\onlinecite{pallecchi2002:PRB}. Using the enhanced $g$-factor is consistent with the fact that the tunneling in our model is dominated by the lowest transverse channel.

Reference \onlinecite{amasha2008:PRB} showed a strong variation in $\chi$ as the potentials defining the confinement of the dot were changed. It was concluded that $\chi$ strongly depends on the dot's shape. Our theory does not account for such a dependence --- the data in Fig.~\ref{fig:g-factor}d show only a marginal dependence of $\chi$ on $l_x-l_y$. 
There are two mechanisms by which changing the gate voltages defining the dot confinement may influence the tunneling asymmetry. First, changes in those voltages can be accompanied by shifts in the position of the dot and thus modulations of the barrier length $L$ and of the tunneling asymmetry.~\cite{footnote4} Second, shape voltages can directly influence the transverse confinement in the lead tip, on which the $g$-factor strongly depends.~\cite{koop2007:JSNM}

\section{Eigenspinor orientation mismatch}

We next investigate the influence that spin-orbit interactions have on $\chi$, assuming a homogeneous $g$-factor. The numerical results allow us to rule out the spin-orbit interactions as the source for the observed asymmetry, as the effect is too small. For the sake of completeness, we nevertheless explain how the spin-dependent tunneling arises, since it may become important in materials with a stronger spin-orbit coupling.

It is important to note that the effective strength of spin-orbit interactions is not
the same in the dot and in the lead. In 2DEG GaAs, the spin-orbit induced magnetic field is $\sim$10 Tesla. As this is at least comparable to the external magnetic field, spin-orbit interactions significantly influence the eigenspinor orientation. On the other hand, spin-orbit interactions are strongly suppressed in the dot. The spin is almost perfectly aligned with the external magnetic field, deflecting from it by a small angle of the order of $l_{x,y}/l_{\rm so} \ll 1$.\cite{aleiner2001:PRL,levitov2003:PRB} Due to the misaligned eigenspinors, 
spin-dependent tunneling is expected.

\begin{figure}
\centerline{\psfig{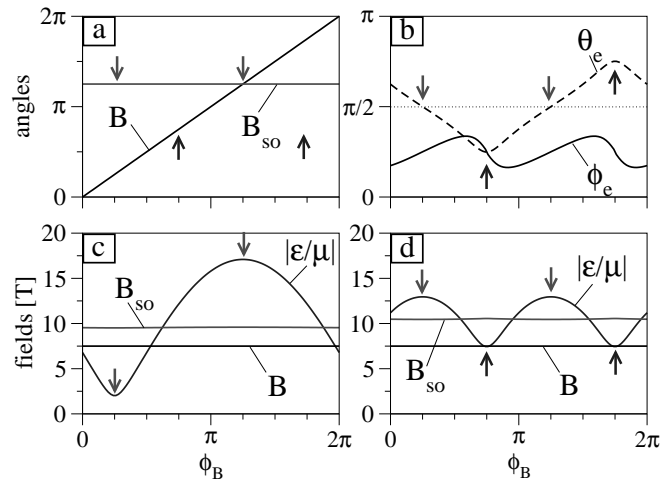}}
\caption{Description of extended and evanescent spinors. (a) In-plane direction of the external and spin-orbit magnetic field. At a downward/upward arrow the fields are (anti)parallel/perpendicular. (b) Azimuthal ($\phi$) and inclination ($\theta$) angles for the spin direction of an evanescent spinor. [(c)-(d)] External and spin-orbit magnetic fields, and Zeeman energy rescaled by $\mu=(g/2)\mu_B$ for (c) extended and (d) evanescent spinor.}
\label{fig:so-waves}
\end{figure}

We therefore consider the eigenfunctions of the Hamiltonian in Eq.~(\ref{eq:Hamiltonian}) for a constant potential $V({\bf r})=V$. The ansatz\cite{sablikov2007:PRB} $\Psi({\bf r}) = \exp( {\rm i}{\bf k} \cdot {\bf r}) \xi$ leads to an algebraic equation for the wavevector ${\bf k}$ and a spatially independent two component eigenspinor $\xi$, 
\begin{equation}
\left\{\frac{\hbar {\bf k}^2}{2 m } + V - E + \frac{g}{2}\mu_B [{\bf B} + {\bf B_{\rm so}(k)}] \cdot \boldsymbol{\sigma} \right\} \xi=0.
\label{eq:spin in complex field}
\end{equation}
For given energy and wavevector, Eq.~(\ref{eq:spin in complex field}) has two solutions, which we label by the spin index $\sigma=\pm1$. The direction of the eigenspinor $\xi_\sigma$ is parametrized by its inclination and azimuthal angles $(\theta_\sigma, \phi_\sigma)$. We call a state evanescent if $V>E$ and extended if $V<E$. These two have purely imaginary and real wavevector ${\bf k}$, respectively, if one neglects the Zeeman energy $\epsilon_\sigma$ contribution to the eigenstate total energy $E$. The spin-orbit induced magnetic field ${\bf B_{\rm so}}$, defined by Eq.~(\ref{eq:spin-orbit}), is in-plane, and generates an anisotropy if the external field direction is varied with respect to the crystallographic axes. In Fig.~\ref{fig:so-waves}a, we plot the azimuthal angle  of ${\bf B}$ and ${\bf B_{\rm so}}$ and mark the directions where the external field is (anti)parallel/perpendicular with the spin-orbit field by the downward/upward arrows. 

For an extended spinor, ${\bf B_{\rm so}(k)}$ is real. The sum of the spin-orbit and external fields sets the spin direction of the extended solutions of Eq.~(\ref{eq:spin in complex field}) to
\begin{equation}
\theta_\sigma=\pi/2, \, \phi_\sigma=\phi_{{\bf B}+{\bf B_{so}}}+(1-\sigma)\pi/2, \, \epsilon_\sigma = \sigma \epsilon.
\end{equation}
The Zeeman splitting energy
\begin{equation}
\epsilon = (g/2)\mu_B \sqrt{({\bf B}+{\bf B_{\rm so}})^2}
\label{eq:Zeeman energy} 
\end{equation}
is (minimal) maximal, if the two fields are (anti)parallel. This is shown in Fig.~\ref{fig:so-waves}c. The opposite Zeeman energy of the two spinors is compensated by a small change in the wavevectors.

The picture changes appreciably for an evanescent spinor. Despite their importance for tunneling, analytical solutions of Eq.~(\ref{eq:spin in complex field}) are known only for zero magnetic field,\cite{sablikov2007:PRB} while some numerical results exist for a finite magnetic field.\cite{lee2005:PRB, usaj2005:EL, serra2007:PRB} A simple perturbative solution of Eq.~(\ref{eq:spin in complex field}) follows if the spin-orbit magnetic field is taken as
\begin{equation}
{\bf B_{\rm so}(k)} \approx {\bf B_{\rm so}(k_0)}, 
\label{eq:bso approximation}
\end{equation}
with the unperturbed wavevector defined by $\hbar {\bf k}_{\bf 0}^2 / 2 m = E - V<0$. Expanding the square root in Eq.~(\ref{eq:Zeeman energy}), one finds that the error of the approximation is of order $(E_{\rm so}/|E - V|)^2 \sim 10^{-4}$. The spin-orbit magnetic field is now purely imaginary. Still, the Zeeman interaction term  with a complex ``magnetic field''\cite{nguyen2009:PRB} has two eigenstates with opposite complex eigenenergies given by Eq.~(\ref{eq:Zeeman energy}), whose magnitude is plotted in Fig.~\ref{fig:so-waves}d. In the coordinate system with the $z$ axis perpendicular to the plane defined by ${\bf B}$ and ${\bf B_{\rm so}}$ (the crystallographic coordinate system here), the spin direction of the evanescent eigenspinors is, 
\begin{equation}
\theta_\sigma = \theta_e, \, \phi_\sigma=\phi_e+(1-\sigma)\pi/2, \, \epsilon_\sigma = \sigma \epsilon.
\end{equation}
The angles $\theta_e$, and $\phi_e$ are plotted in Fig.~\ref{fig:so-waves}b as a function of the external magnetic field orientation. The two spinors are orthogonal and in-plane only if the two magnetic fields are (anti)parallel (downward arrows). The out-of-plane spinor component is maximal if the two fields are orthogonal (upward arrows). 

With the Zeeman energy given by Eqs.~(\ref{eq:Zeeman energy}) and (\ref{eq:bso approximation}), the wavevector ${\bf k}$ is obtained from the quadratic Eq.~(\ref{eq:spin in complex field}). The imaginary part of the Zeeman term and the kinetic energy cancel such that the total energy $E$ is real. To illustrate the solutions, we consider the wavevector fixed along ${\bf k_0}$ so we can write it as ${\bf k}=(1-\kappa_{\rm z}/k_0-{\rm i}\kappa_{\rm so}/k_0){\bf k_0}$ introducing two real parameters $\kappa_{\rm z/so}$. To leading order, these are given by
\begin{equation}
\kappa_{\rm z} + {\rm i} \kappa_{\rm so} = \pm \frac{2m \epsilon}{\hbar^2 k_0}, 
\end{equation}
in particular, they are opposite for the two spinors. 

Consider now that a coherent superposition of the two eigenspinors propagates inside the barrier. The real part of the Zeeman energy leads to different decay lengths for the two components, while the imaginary part leads to a coherent rotation of the initial spin orientation. If both the magnetic field and the spin-orbit interactions are present, the spin will make a spiral, both decaying and rotating at the same time, over the length scales $1/\kappa_{\rm z}$, and $1/\kappa_{\rm so}$, respectively. The evanescent states show an additional possibility for the spin selection through spin dependent penetration lengths. Note, however, that both eigenspinors are additionally damped over the length $1/k_0$, which is usually much shorter than $1/\kappa_{\rm z,so}$, strongly limiting the achievable differentiation for the two spin species. 

Before we present numerical results we note an interesting singular limit when the external and the spin-orbit magnetic fields are orthogonal and of the same magnitude. In this case the out-of-plane spinor misalignment is maximal and both spinors are oriented along the same direction (the $z$ axis).

\begin{figure}
\centerline{\psfig{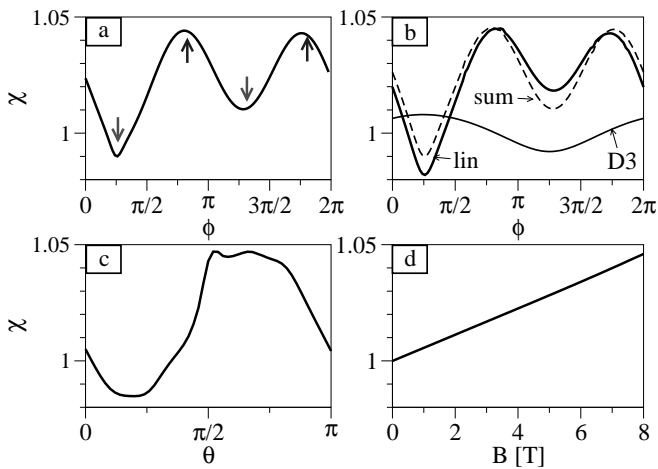}}
\caption{Influence of the magnetic field strength and orientation on the spin-orbit induced spin tunneling asymmetry. (a) At a  downward/upward arrow the external and spin-orbit magnetic fields are (anti)parallel/perpendicular. (b) The asymmetry if only linear (lin), cubic (D3), or both (dashed) spin-orbit interactions are present. (c) Asymmetry for varied out-of-plane field orientation. This is the only case in this manuscript, where the vector potential ${\bf A}$ contributes to the kinematic momentum operator, $\hbar {\bf k}=-{\rm i}\hbar \boldsymbol{\nabla}+e{\bf A}$. (d) Asymmetry as a function of the in-plane field strength.}
\label{fig:magnetic field}
\end{figure}

We analyze our numerical results for the tunneling asymmetry using the just developed picture of the extended and evanescent spinors. In Fig.~\ref{fig:magnetic field}a we vary the direction of the external magnetic field. If the two magnetic fields are (anti)parallel (downward arrows), the spin quantization axis throughout the structure is the same and only spin-diagonal tunneling is possible. The asymmetry arises from the spin-orbit contribution to the Zeeman energy in the lead, which gives rise to spin dependent discrepancies in the wavevector. When the magnetic fields are not parallel, the eigenspinor orientations in the lead, under the barrier and in the dot are all different, which opens off-diagonal spin tunneling channels. The spin dependent tunneling arises from different effective potential barriers in these off-diagonal channels. The wavevectors in Eq.~(\ref{eq:wavevector}) fulfill [note that $\sigma\to\sigma^\prime$ refers to an eigenspinor direction in the lead ($\sigma$) and the dot ($\sigma^\prime$), which are not necessarily along the same axis]
\begin{equation}
 k_{\downarrow \to \uparrow} > k_{\uparrow \to \uparrow} = k_{\downarrow\to\downarrow} > k_{\uparrow\to\downarrow}. 
\label{eq:wavevectors}
\end{equation}
They are shown as arrows of different lengths in Fig.~\ref{fig:mechanisms}e. The tunneling into the excited state is always preferred and the effect is maximal if the spin-orbit and the magnetic fields in the lead are perpendicular. 

Figure \ref{fig:magnetic field}b shows that the asymmetry is dominated by the linear spin-orbit terms, with the cubic Dresselhaus contribution smaller by a factor of $\sim$5. We checked (but do not show) that the contributions from different spin-orbit terms are additive with good accuracy. In Fig.~\ref{fig:magnetic field}c we consider an out-of-plane magnetic field and show that, despite the presence of orbital effects, the asymmetry is of similar magnitude as for an in-plane field. Finally, we show in Fig.~\ref{fig:magnetic field}d that the asymmetry grows with the magnetic field. This is expected, as a larger Zeeman energy makes the difference of wavevectors in Eq.~(\ref{eq:wavevectors}) larger. 

\begin{figure}
\centerline{\psfig{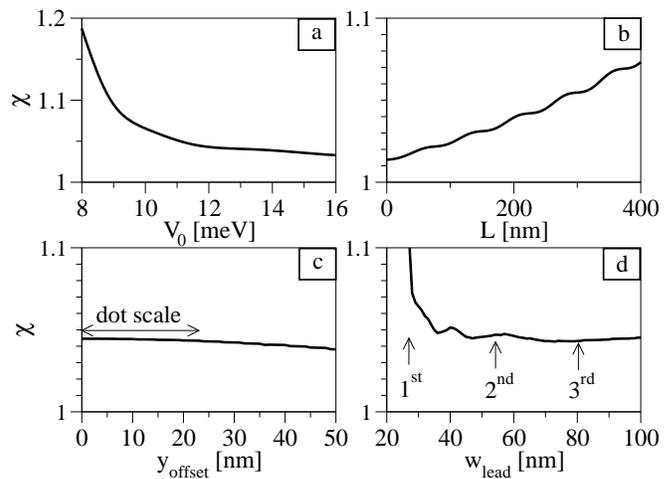}}
\caption{Dependence of the spin-orbit induced spin tunneling asymmetry on (a) the barrier height $V_0$, (b) the barrier ramp $L$ [see Fig.~\ref{fig:mechanisms}a], (c) the dot-lead offset $y_{\rm offset}$ [see Figs.~\ref{fig:mechanisms}a and \ref{fig:mechanisms}b], (d) the lead width $w_{\rm width}$. The widths at which the lowest three transverse channels become open are indicated. In all panels, the tunneling rate into the ground state has been kept fixed at 200 Hz.}
\label{fig:barrier}
\end{figure}

Figure \ref{fig:barrier} shows the dependence of the effect on the barrier potential. The tunneling asymmetry is more pronounced for lower (panel a) and longer (panel b) barriers because the longer the barrier, the stronger the suppression of the spin down-to-up tunneling channel relative to the up-to-down channel. We next see in Fig.\ref{fig:barrier}c that the dot's symmetry has little importance since varying the dot-lead offset changes the tunneling asymmetry only marginally. Similarly, Fig.~\ref{fig:barrier}d shows that the lead width is important only around the pinch-off point, where the upper spin state disappears from the lead. Because the tunneling is dominated by the lowest transverse channel contribution, opening additional channels has a hardly noticeable effect on $\chi$.

Our numerical data therefore show that spin-orbit induced spin tunneling asymmetries reflect evanescent and extended spinor solutions of Eq.~(\ref{eq:spin in complex field}) but that the effect is too small to account for experimental observations. Here also, the dot symmetry plays only a minor role.

\section{Conclusions}

We investigated possible spin dependencies in the tunneling of an electron into an empty lateral quantum dot in presence of a magnetic field. We used a general two dimensional model and examined the parameters' space in great details, by varying the tunnel barrier geometry, the spin-orbit interactions strengths, the magnetic field orientation and strength, the dot confinement potential, the dot $g$-factor, and the dot vs lead orientation. We found that the spin tunneling asymmetry does not exceed 10\% for realistic parameters, much too small compared to the results of Ref.~\onlinecite{amasha2008:PRB}. We therefore postulate that the observed asymmetry originates in an anomalously enhanced lead $g$-factor, because of the lateral confinement in the lead. Such an enhancement has been reported in Refs.~\onlinecite{thomas1996:PRL} and \onlinecite{chen2009:PRB}.

We expect the error stemming from approximations in our model to be negligible compared to the specificities of the tunnel barrier profile, about which not much is known in gated quantum dots. On the other hand, we do not expect the asymmetry to arise because of specific barrier shape, as seemingly related effects are present in different structures.\cite{potok2003:PRL} 

Our findings provide cues for further experimental checks of the origin of the spin tunneling asymmetry. Assuming the $g$-factor mechanism is at play, the tunneling asymmetry growth is accompanied by a reduction in the tunneling rates themselfs. On the other hand, the spin-orbit interactions induce anisotropy with respect to the crystallographic directions and we identify specific directions of the magnetic field, which give maximal/minimal spin tunneling asymmetry. This mechanism might be of relevance in other materials, such as InAs.\cite{meier2007:NP} We note, but do not show, that our numerics indicate that these two effects are essentially additive (independent).

Among alternative explanations for the observed asymmetry, one could consider that single electron tunneling is dynamically assisted by phonons and/or nuclear spins. However, one does not expect these to induce a spin-dependence, as phonons do not couple and nuclear spins couple only very weakly to the electronic spin. The dependence may arise due to an inhomogeneous collective nuclear field, which would be similar to, and presumably indistinguishable from, the inhomogeneous $g$-factor that we studied here. We note finally, that the finite extension of the 2DEG in the $z$-direction allows for orbital effects. The fields of up to 7-8 Tesla used in Ref.~\onlinecite{amasha2008:PRB} however correspond to a magnetic length significantly larger than the typical thickness of the 2DEG . Possible weak diamagnetic effects, such as a slight shift or a compression of the 2DEG, seem to affect both the lead and the dot in the same way and should not result in the spin tunneling asymmetry.

\acknowledgements

We would like to thank Dominik Zumb\"uhl for valuable discussions and reminding us of the $g$-factor enhancement in Ref.~\onlinecite{thomas1996:PRL}, and Sami Amasha and Kenneth MacLean for a detailed discussion of Ref.~\onlinecite{amasha2008:PRB} and of related unpublished experimental data. This work has been supported by the NSF under Grant No.~ DMR-0706319.

\appendix

\section{Comparison of various formulas for tunneling}

We compare four different formulas for the dot to lead tunneling rate in one dimension. We differentiate scattering approaches, based on eigenstates of the full system Hamiltonian from perturbative approaches, where the dot and the lead are considered separately. We find that the perturbative approach used in the main text agrees very well with scattering approaches. A similar correspondence is expected for tunneling into a quasi one-dimensional channels of a finite-width lead which we consider in the main text. 

\begin{figure}
\centerline{\psfig{file=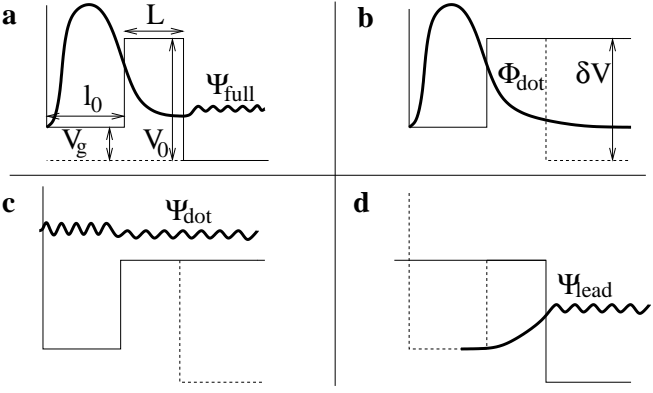,width=1\linewidth}}
\caption{(a) Full scattering eigenstate (thick line) and the confinement potential (thin solid line) of the dot-barrier-lead system. (b) Potential of an isolated dot and its bound eigenstate and the perturbation $\delta V$. (c) Extended eigenstate for the same problem. (d) Potential profile and an extended eigenstate of an isolated lead.}
\label{fig:tunneling cartoon}
\end{figure}

We set up a one dimensional square barrier model with the Hamiltonian $H=\hbar^2 k^2/2m + V(x)$. The potential $V(x)$, depicted in Fig.~\ref{fig:tunneling cartoon}a, describes a dot of width $l_0$, separated from a semi-infinite lead by a barrier of width $L$. The dot is offset with respect to the lead by $V_g$ and the barrier height is $V_0$. We compare the rates for tunneling to all quasi-bound states of the structure. 

We take the Hilbert space to be spanned by a bound state $|\phi_0\rangle$ and a set of (orthonormal) extended states $\{ |\psi_n\rangle \}_{n \neq 0}$. We expand the time dependent system wavefunction as\cite{gurvitz1987:PRL} 
\begin{equation}
|\Psi\rangle = e^{-{\rm i} \omega_0 t} a_0 |\phi_0\rangle + \sum_n  e^{-{\rm i} \omega_n t} a_n |\psi_n\rangle,
\label{eq:basis}
\end{equation}
introducing so far unspecified frequencies $\omega_i$ and time dependent coefficients $a_i(t)$. The time dependent Schr\"odinger equation for $|\Psi\rangle$ reads
\begin{subequations}
\begin{eqnarray}
{\rm i}\hbar \partial_t a_0 &=& - \langle \phi_0 | \hbar \omega_0 - H | \phi_0 \rangle a_0
\label{eq:schrodingera}\\ 
\nonumber 
&&-\sum_n e^{-{\rm i} \omega_{n0} t} \langle \phi_0 | {\rm i}\hbar \partial_t + \hbar \omega_n - H | \psi_n \rangle a_n,\\
{\rm i}\hbar \partial_t a_n &=& - e^{{\rm i} \omega_{n0} t} \langle \psi_n | {\rm i}\hbar \partial_t + \hbar \omega_0 - H | \phi_0 \rangle a_0
\label{eq:schrodingerb}\\ 
\nonumber 
&&- \sum_m e^{-{\rm i} \omega_{mn} t} \langle \psi_n | \hbar \omega_m - H | \psi_m \rangle a_m,
\end{eqnarray}
\label{eq:schrodinger}
\end{subequations}
where we denote $\omega_{ij}=\omega_i-\omega_j$.

Let us now consider the lead as a perturbation of an isolated dot, as shown in Fig.~\ref{fig:tunneling cartoon}b, that is, $H=H_0+\delta \hat{V}$, where
\begin{equation}
\delta \hat{V} = - \theta(x-l_0-L) \delta V \equiv - \theta(x-l_0-L) V_0.
\end{equation}
Accordingly, we choose $|\phi_0\rangle$ and $|\psi_n\rangle\equiv|\psi_n^{\rm dot}\rangle$ as the bound and extended eigenstates of $H_0$, depicted in Figs.~\ref{fig:tunneling cartoon}b and \ref{fig:tunneling cartoon}c, respectively, and $\hbar \omega_i$ their energies. The formula for the tunneling rate follows from Eq.~(\ref{eq:schrodinger}) as\cite{peres1980:JPA}
\begin{equation}
\Gamma_1=\frac{2\pi}{\hbar} \sum_n |\langle \phi_0 | \delta \hat{V} | \psi^{\rm dot}_n \rangle |^2 
\delta(\hbar \omega_0 - \hbar \omega_n^{\rm dot}+\delta V),
\label{eq:FGR1}
\end{equation}
if one neglects the transitions between extended states [$m\neq n$ terms in Eq.~(\ref{eq:schrodingerb})] and the bound state energy shift, i.e. $\langle \phi_0 | \delta \hat{V} | \phi_0 \rangle \approx 0$.

Equation (\ref{eq:FGR1}) turns out to be a poor estimate. Gurvitz {\it et al.~}\cite{gurvitz1987:PRL,gurvitz1988:PRA} suggested to improve it by replacing the extended states of the isolated dot $|\psi_n^{\rm dot}\rangle$ [see Fig.~\ref{fig:tunneling cartoon}c] by the extended states of the isolated lead $|\psi_n^{\rm lead}\rangle$ [see Fig.~\ref{fig:tunneling cartoon}d]. We obtain Eq.~(\ref{eq:fermis}),
\begin{equation}
\Gamma_2 = \frac{2\pi}{\hbar} \sum_n |\langle \phi_0 | \delta \hat{V} | \psi^{\rm lead}_n \rangle |^2 
\delta(\hbar \omega_0 - \hbar \omega_n^{\rm lead}).
\label{eq:FGR2}
\end{equation}
Equation (\ref{eq:FGR2}) is equivalent to the Bardeen formula for tunneling,\cite{bardeen1961:PRL} which comes from the following derivation. In Eq.~(\ref{eq:basis}), we choose $|\phi_0\rangle$ and $|\psi_n\rangle$ to be the dot bound state [Fig.~\ref{fig:tunneling cartoon}b] and lead extended states [Fig.~\ref{fig:tunneling cartoon}d]. Transitions stem from those states being neither orthogonal nor eigenstates of $H$. Neglecting continuum to continuum transitions again in Eq.~(\ref{eq:schrodingerb}) and the time derivatives of the coefficients $a_i$ on the right hand side of Eq.~(\ref{eq:schrodinger}) and using that for $\omega_0=\omega_n$ we can write 
\begin{equation}
\langle \phi_0 | H - \hbar \omega_n | \psi_n \rangle = \langle \phi_0 | H - \hbar \omega_0 | \psi_n \rangle = \langle \phi_0 | \delta \hat{V} | \psi_n \rangle.
\end{equation}
We then get Eq.~(\ref{eq:FGR2}) from Eq.~(\ref{eq:schrodinger}).

\begin{figure}
\centerline{\psfig{file=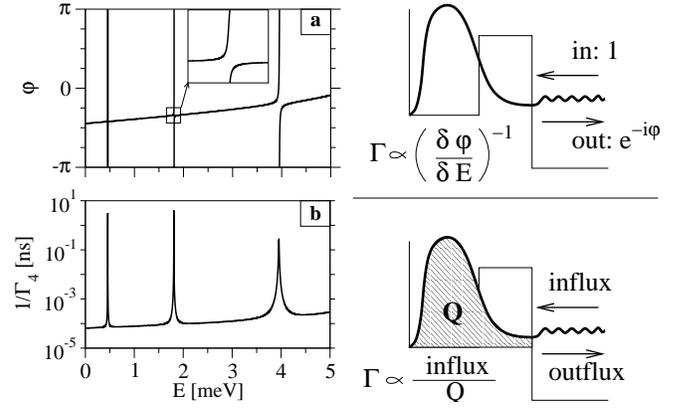,width=1\linewidth}}
\caption{Comparison of two scattering approaches. (a) Wigner method: the reflection phase with a magnified resonance. (b) The flux method: lifetime according to Eq.~(\ref{eq:jalabert}). The methods' schematics are on the right. The model parameters [depicted in Fig.~\ref{fig:tunneling cartoon}a] are: $V_g=7$ meV, $V_0=13$ meV, $L=50$ nm, and $l_0=100$ nm.}
\label{fig:comparison2}
\end{figure}

We next turn our attention to scattering approaches. The first one is due to Wigner.\cite{wigner1955:PR} We write a scattering state (an unnormalized eigenstate of $H$) in the lead as
\begin{equation}
\langle x_{\rm lead}|\psi_{\rm full}\rangle = e^{-{\rm i} k x_{\rm lead}} + e^{{\rm i} k x_{\rm lead}-{\rm i}\varphi}.
\label{eq:incoming reflected}
\end{equation}
Current conservation requires that the reflection results only in a phase shift $\varphi$. Consider now a wave packet of energy $E$ and wavevector $-k$ (velocity $v=\hbar k/m$), scattering at the barrier from the right. The reflected wave phase shift in Eq.~(\ref{eq:incoming reflected}) is interpreted as an action phase due to time of the tunneling $\varphi=Et/\hbar=2 E \Gamma^{-1} /\hbar$. Differentiating with respect to the energy we get
\begin{equation}
\frac{\partial \varphi}{\partial E} =  \frac{2E}{\hbar}\frac{\partial \Gamma^{-1} }{\partial E} + \frac{2\Gamma^{-1}}{\hbar}.
\end{equation}
The quasi-bound state lifetime $\Gamma^{-1}$ is sharply peaked around the resonance energy, where its first derivative is zero. The tunneling rate follows as 
\begin{equation}
\Gamma_3  =  \frac{2}{\hbar} \left( \frac{\partial \varphi}{\partial E} \right)^{-1}.
\label{eq:wigner}
\end{equation}
The reflection phase $\varphi$ is plotted in Fig.~\ref{fig:comparison2}a. The sharp resonances where the phase jumps by 2$\pi$ are evident.

\begin{figure}
\centerline{\psfig{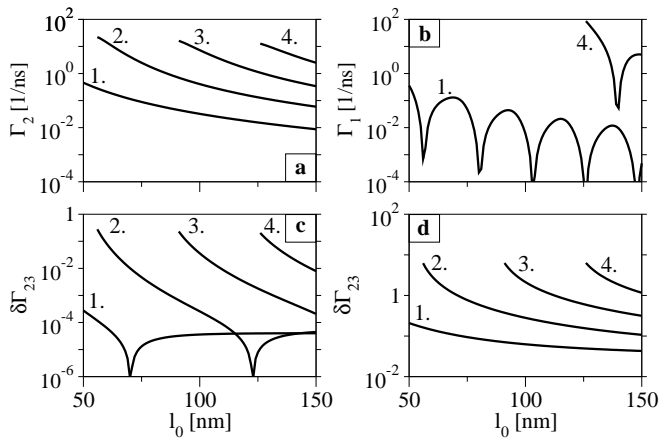}}
\caption{Comparison of tunneling formulas varying the width of the potential well $l_0$. Parameters are as in Fig.~\ref{fig:comparison2}. Tunneling rate for the dot bound states computed using (a) Eq.~(\ref{eq:FGR2}) and (b) Eq.~(\ref{eq:FGR1}), where only two states are shown for clarity.
(c) Comparison of Eqs.~(\ref{eq:FGR2}) and (\ref{eq:wigner}).
(d) Comparison of Eqs.~(\ref{eq:FGR2}) and (\ref{eq:wigner}), with the latter evaluated at the energy of the isolated dot state. In (c) and (d), the relative difference $\delta \Gamma_{ij}=|\Gamma_i-\Gamma_j|/\Gamma_i$ is plotted.}
\label{fig:comparison}
\end{figure}

The second scattering approach we consider is the following. Take a steady state where a constant flux impinges on the barrier. A part of it is directly reflected while a part tunnels through. The latter (influx) increases the probability accumulated behind the barrier, compensated by a constant leak out (outflux). The tunneling rate can be defined as the ratio of the influx and the accumulated probability,\cite{jalabert1992:PRL}
\begin{equation}
\Gamma_4 = \frac{j_{\rm in}}{Q} = \frac{v \alpha |\Psi_{\rm full}^{\rm leftgoing}(x_{\rm lead}=0)|^2}{\int_{\rm dot+barrier} |\Psi_{\rm full}(x)|^2 {\rm d} x}.
\label{eq:jalabert}
\end{equation}
The coefficient $\alpha$ quantifies how much of the left-going probability flux tunnels through (a part is reflected right at the barrier). In analogy with an infinite barrier, we suppose the phase of a reflected wave changes sign at the reflection point, which gives $\alpha = \cos^2 (\varphi/2)$. As on resonance $\varphi=\pi/2$, we get $\alpha=1/2$. Figure \ref{fig:comparison2}b shows the quasi-bound state lifetime computed by Eq.~(\ref{eq:jalabert}). The resonant energies correspond to those obtained in the Wigner method.

We compare the results of the discussed approaches in Fig.~\ref{fig:comparison}. We take the Bardeen-Gurvitz formula, Eq.~(\ref{eq:FGR2}), as reference and plot the corresponding rates in Fig.~\ref{fig:comparison}a. The Fermi Golden rule of Eq.~(\ref{eq:FGR1}) shows spurious resonances and can differ by orders of magnitude, as shown in Fig.~\ref{fig:comparison}b. As $\Gamma_3=\Gamma_4$ within numerical precision of our software, we compare only the Wigner formula to the reference in Fig.~\ref{fig:comparison}c and find an excellent agreement even improving as the resonance energy is reduced. 

To investigate the spin tunneling asymmetry, we have chosen to use the Bardeen-Gurvitz formula, Eq.~(\ref{eq:FGR2}). Out of those we discussed, it presents the advantage that it is based on the eigenstates of the isolated subsystems, which are much easier to obtain than the scattering eigenstates of the full system. An additional disadvantage of the scattering approaches is that as the tunneling rate drops, the resonances become narrower and harder to spot. Therefore, instead of searching for the them numerically, one can take the energy of a dot bound state as the approximation for the resonance position. In Fig.~\ref{fig:comparison2}d we evaluate the rate in the Wigner approach at such approximate resonance energy. For states well below the barrier, we still find a very good correspondence with the Bardeen-Gurvitz formula.

\bibliography{../../references/quantum_dot}

\begin{thebibliography}{46}
\expandafter\ifx\csname natexlab\endcsname\relax\def\natexlab#1{#1}\fi
\expandafter\ifx\csname bibnamefont\endcsname\relax
  \def\bibnamefont#1{#1}\fi
\expandafter\ifx\csname bibfnamefont\endcsname\relax
  \def\bibfnamefont#1{#1}\fi
\expandafter\ifx\csname citenamefont\endcsname\relax
  \def\citenamefont#1{#1}\fi
\expandafter\ifx\csname url\endcsname\relax
  \def\url#1{\texttt{#1}}\fi
\expandafter\ifx\csname urlprefix\endcsname\relax\def\urlprefix{URL }\fi
\providecommand{\bibinfo}[2]{#2}
\providecommand{\eprint}[2][]{\url{#2}}

\bibitem[{\citenamefont{Fert}(2008)}]{fert2008:RMP}
\bibinfo{author}{\bibfnamefont{A.}~\bibnamefont{Fert}}, \bibinfo{journal}{Rev.
  Mod. Phys.} \textbf{\bibinfo{volume}{80}}, \bibinfo{pages}{1517}
  (\bibinfo{year}{2008}).

\bibitem[{\citenamefont{{Gr\"unberg}}(2008)}]{grunberg2008:RMP}
\bibinfo{author}{\bibfnamefont{A.}~\bibnamefont{{Gr\"unberg}}},
  \bibinfo{journal}{Rev. Mod. Phys} \textbf{\bibinfo{volume}{80}},
  \bibinfo{pages}{1531} (\bibinfo{year}{2008}).

\bibitem[{\citenamefont{Fabian et~al.}(2007)\citenamefont{Fabian,
  {Matos-Abiague}, Ertler, Stano, and {\v{Z}uti\'{c}}}}]{fabian2007:APS}
\bibinfo{author}{\bibfnamefont{J.}~\bibnamefont{Fabian}},
  \bibinfo{author}{\bibfnamefont{A.}~\bibnamefont{{Matos-Abiague}}},
  \bibinfo{author}{\bibfnamefont{C.}~\bibnamefont{Ertler}},
  \bibinfo{author}{\bibfnamefont{P.}~\bibnamefont{Stano}}, \bibnamefont{and}
  \bibinfo{author}{\bibfnamefont{I.}~\bibnamefont{{\v{Z}uti\'{c}}}},
  \bibinfo{journal}{Acta Phys. Slov.} \textbf{\bibinfo{volume}{57}},
  \bibinfo{pages}{565} (\bibinfo{year}{2007}).

\bibitem[{\citenamefont{Hanson et~al.}(2007)\citenamefont{Hanson, Kouwenhoven,
  Petta, Tarucha, and Vandersypen}}]{hanson2007:RMP}
\bibinfo{author}{\bibfnamefont{R.}~\bibnamefont{Hanson}},
  \bibinfo{author}{\bibfnamefont{L.~P.} \bibnamefont{Kouwenhoven}},
  \bibinfo{author}{\bibfnamefont{J.~R.} \bibnamefont{Petta}},
  \bibinfo{author}{\bibfnamefont{S.}~\bibnamefont{Tarucha}}, \bibnamefont{and}
  \bibinfo{author}{\bibfnamefont{L.~M.~K.} \bibnamefont{Vandersypen}},
  \bibinfo{journal}{Rev. Mod. Phys.} \textbf{\bibinfo{volume}{79}},
  \bibinfo{pages}{1217} (\bibinfo{year}{2007}).

\bibitem[{\citenamefont{Taylor et~al.}(2007)\citenamefont{Taylor, Petta,
  Johnson, Yacoby, Marcus, and Lukin}}]{taylor2007:PRB}
\bibinfo{author}{\bibfnamefont{J.~M.} \bibnamefont{Taylor}},
  \bibinfo{author}{\bibfnamefont{J.~R.} \bibnamefont{Petta}},
  \bibinfo{author}{\bibfnamefont{A.~C.} \bibnamefont{Johnson}},
  \bibinfo{author}{\bibfnamefont{A.}~\bibnamefont{Yacoby}},
  \bibinfo{author}{\bibfnamefont{C.~M.} \bibnamefont{Marcus}},
  \bibnamefont{and} \bibinfo{author}{\bibfnamefont{M.~D.} \bibnamefont{Lukin}},
  \bibinfo{journal}{Phys. Rev. B} \textbf{\bibinfo{volume}{76}},
  \bibinfo{pages}{035315} (\bibinfo{year}{2007}).

\bibitem[{\citenamefont{Elzerman et~al.}(2004)\citenamefont{Elzerman, Hanson,
  {van Beveren}, Witkamp, Vandersypen, and Kouwenhoven}}]{elzerman2004:N}
\bibinfo{author}{\bibfnamefont{J.~M.} \bibnamefont{Elzerman}},
  \bibinfo{author}{\bibfnamefont{R.}~\bibnamefont{Hanson}},
  \bibinfo{author}{\bibfnamefont{L.~H.~W.} \bibnamefont{{van Beveren}}},
  \bibinfo{author}{\bibfnamefont{B.}~\bibnamefont{Witkamp}},
  \bibinfo{author}{\bibfnamefont{L.~M.~K.} \bibnamefont{Vandersypen}},
  \bibnamefont{and} \bibinfo{author}{\bibfnamefont{L.~P.}
  \bibnamefont{Kouwenhoven}}, \bibinfo{journal}{Nature}
  \textbf{\bibinfo{volume}{430}}, \bibinfo{pages}{431} (\bibinfo{year}{2004}).

\bibitem[{\citenamefont{Pfund et~al.}(2009)\citenamefont{Pfund, Shorubalko,
  Ensslin, and Leturcq}}]{pfund2009:PRB}
\bibinfo{author}{\bibfnamefont{A.}~\bibnamefont{Pfund}},
  \bibinfo{author}{\bibfnamefont{I.}~\bibnamefont{Shorubalko}},
  \bibinfo{author}{\bibfnamefont{K.}~\bibnamefont{Ensslin}}, \bibnamefont{and}
  \bibinfo{author}{\bibfnamefont{R.}~\bibnamefont{Leturcq}},
  \bibinfo{journal}{Phys. Rev. B} \textbf{\bibinfo{volume}{79}},
  \bibinfo{pages}{121306(R)} (\bibinfo{year}{2009}).

\bibitem[{\citenamefont{Hanson et~al.}(2005)\citenamefont{Hanson, {van
  Beveren}, Vink, Elzerman, Naber, Koppens, Kouwenhoven, and
  Vandersypen}}]{hanson2005:PRL}
\bibinfo{author}{\bibfnamefont{R.}~\bibnamefont{Hanson}},
  \bibinfo{author}{\bibfnamefont{L.~H.~W.} \bibnamefont{{van Beveren}}},
  \bibinfo{author}{\bibfnamefont{I.~T.} \bibnamefont{Vink}},
  \bibinfo{author}{\bibfnamefont{J.~M.} \bibnamefont{Elzerman}},
  \bibinfo{author}{\bibfnamefont{W.~J.~M.} \bibnamefont{Naber}},
  \bibinfo{author}{\bibfnamefont{F.~H.~L.} \bibnamefont{Koppens}},
  \bibinfo{author}{\bibfnamefont{L.~P.} \bibnamefont{Kouwenhoven}},
  \bibnamefont{and} \bibinfo{author}{\bibfnamefont{L.~M.~K.}
  \bibnamefont{Vandersypen}}, \bibinfo{journal}{Phys. Rev. Lett.}
  \textbf{\bibinfo{volume}{94}}, \bibinfo{pages}{196802}
  (\bibinfo{year}{2005}).

\bibitem[{\citenamefont{Amasha et~al.}(2008{\natexlab{a}})\citenamefont{Amasha,
  MacLean, Radu, {Zumb\"uhl}, Kastner, Hanson, and Gossard}}]{amasha2008:PRB}
\bibinfo{author}{\bibfnamefont{S.}~\bibnamefont{Amasha}},
  \bibinfo{author}{\bibfnamefont{K.}~\bibnamefont{MacLean}},
  \bibinfo{author}{\bibfnamefont{I.~P.} \bibnamefont{Radu}},
  \bibinfo{author}{\bibfnamefont{D.~M.} \bibnamefont{{Zumb\"uhl}}},
  \bibinfo{author}{\bibfnamefont{M.~A.} \bibnamefont{Kastner}},
  \bibinfo{author}{\bibfnamefont{M.~P.} \bibnamefont{Hanson}},
  \bibnamefont{and} \bibinfo{author}{\bibfnamefont{A.~C.}
  \bibnamefont{Gossard}}, \bibinfo{journal}{Phys. Rev. B}
  \textbf{\bibinfo{volume}{78}}, \bibinfo{pages}{041306(R)}
  (\bibinfo{year}{2008}{\natexlab{a}}).

\bibitem[{\citenamefont{Voskoboynikov et~al.}(1999)\citenamefont{Voskoboynikov,
  Liu, and Lee}}]{voskoboynikov1999:PRB}
\bibinfo{author}{\bibfnamefont{A.}~\bibnamefont{Voskoboynikov}},
  \bibinfo{author}{\bibfnamefont{S.~S.} \bibnamefont{Liu}}, \bibnamefont{and}
  \bibinfo{author}{\bibfnamefont{C.~P.} \bibnamefont{Lee}},
  \bibinfo{journal}{Phys. Rev. B} \textbf{\bibinfo{volume}{59}},
  \bibinfo{pages}{12514} (\bibinfo{year}{1999}).

\bibitem[{\citenamefont{Voskoboynikov et~al.}(2000)\citenamefont{Voskoboynikov,
  Lin, Lee, and Tretyak}}]{voskoboynikov2000:JAP}
\bibinfo{author}{\bibfnamefont{A.}~\bibnamefont{Voskoboynikov}},
  \bibinfo{author}{\bibfnamefont{S.~S.} \bibnamefont{Lin}},
  \bibinfo{author}{\bibfnamefont{C.~P.} \bibnamefont{Lee}}, \bibnamefont{and}
  \bibinfo{author}{\bibfnamefont{O.}~\bibnamefont{Tretyak}},
  \bibinfo{journal}{J. Appl. Phys.} \textbf{\bibinfo{volume}{87}},
  \bibinfo{pages}{387} (\bibinfo{year}{2000}).

\bibitem[{\citenamefont{Glazov et~al.}(2005)\citenamefont{Glazov, Alekseev,
  Odnoblyudov, Chistyakov, Tarasenko, and Yassievich}}]{glazov2005:PRB}
\bibinfo{author}{\bibfnamefont{M.~M.} \bibnamefont{Glazov}},
  \bibinfo{author}{\bibfnamefont{P.~S.} \bibnamefont{Alekseev}},
  \bibinfo{author}{\bibfnamefont{M.~A.} \bibnamefont{Odnoblyudov}},
  \bibinfo{author}{\bibfnamefont{V.~M.} \bibnamefont{Chistyakov}},
  \bibinfo{author}{\bibfnamefont{S.~A.} \bibnamefont{Tarasenko}},
  \bibnamefont{and} \bibinfo{author}{\bibfnamefont{I.~N.}
  \bibnamefont{Yassievich}}, \bibinfo{journal}{Phys. Rev. B}
  \textbf{\bibinfo{volume}{71}}, \bibinfo{pages}{155313}
  (\bibinfo{year}{2005}).

\bibitem[{\citenamefont{Voskoboynikov et~al.}(1998)\citenamefont{Voskoboynikov,
  Liu, and Lee}}]{voskoboynikov1998:PRB}
\bibinfo{author}{\bibfnamefont{A.}~\bibnamefont{Voskoboynikov}},
  \bibinfo{author}{\bibfnamefont{S.~S.} \bibnamefont{Liu}}, \bibnamefont{and}
  \bibinfo{author}{\bibfnamefont{C.~P.} \bibnamefont{Lee}},
  \bibinfo{journal}{Phys. Rev. B} \textbf{\bibinfo{volume}{58}},
  \bibinfo{pages}{15397} (\bibinfo{year}{1998}).

\bibitem[{\citenamefont{Rozhansky and Averkiev}(2008)}]{rozhansky2008:PRB}
\bibinfo{author}{\bibfnamefont{I.~V.} \bibnamefont{Rozhansky}}
  \bibnamefont{and} \bibinfo{author}{\bibfnamefont{N.~S.}
  \bibnamefont{Averkiev}}, \bibinfo{journal}{Phys. Rev. B}
  \textbf{\bibinfo{volume}{77}}, \bibinfo{pages}{115309}
  (\bibinfo{year}{2008}).

\bibitem[{\citenamefont{Perel et~al.}(2003)\citenamefont{Perel, Tarasenko,
  Yassievich, Ganichev, Belkov, and Prettl}}]{perel2003:PRB}
\bibinfo{author}{\bibfnamefont{V.~I.} \bibnamefont{Perel}},
  \bibinfo{author}{\bibfnamefont{S.~A.} \bibnamefont{Tarasenko}},
  \bibinfo{author}{\bibfnamefont{I.~N.} \bibnamefont{Yassievich}},
  \bibinfo{author}{\bibfnamefont{S.~D.} \bibnamefont{Ganichev}},
  \bibinfo{author}{\bibfnamefont{V.~V.} \bibnamefont{Belkov}},
  \bibnamefont{and} \bibinfo{author}{\bibfnamefont{W.}~\bibnamefont{Prettl}},
  \bibinfo{journal}{Phys. Rev. B} \textbf{\bibinfo{volume}{67}},
  \bibinfo{pages}{201304(R)} (\bibinfo{year}{2003}).

\bibitem[{\citenamefont{St\v{r}eda and \v{S}eba}(2003)}]{streda2003:PRL}
\bibinfo{author}{\bibfnamefont{P.}~\bibnamefont{St\v{r}eda}} \bibnamefont{and}
  \bibinfo{author}{\bibfnamefont{P.}~\bibnamefont{\v{S}eba}},
  \bibinfo{journal}{Phys. Rev. Lett.} \textbf{\bibinfo{volume}{90}},
  \bibinfo{pages}{256601} (\bibinfo{year}{2003}).

\bibitem[{\citenamefont{Silvestrov and Mishchenko}(2006)}]{silvestrov2006:PRB}
\bibinfo{author}{\bibfnamefont{P.~G.} \bibnamefont{Silvestrov}}
  \bibnamefont{and} \bibinfo{author}{\bibfnamefont{E.~G.}
  \bibnamefont{Mishchenko}}, \bibinfo{journal}{Phys. Rev. B}
  \textbf{\bibinfo{volume}{74}}, \bibinfo{pages}{165301}
  (\bibinfo{year}{2006}).

\bibitem[{\citenamefont{Tkach et~al.}(2009)\citenamefont{Tkach, Sablikov, and
  Sukhanov}}]{tkach2009:JPCM}
\bibinfo{author}{\bibfnamefont{Y.~Y.} \bibnamefont{Tkach}},
  \bibinfo{author}{\bibfnamefont{V.~A.} \bibnamefont{Sablikov}},
  \bibnamefont{and} \bibinfo{author}{\bibfnamefont{A.~A.}
  \bibnamefont{Sukhanov}}, \bibinfo{journal}{J. Phys.: Condens. Matter}
  \textbf{\bibinfo{volume}{21}}, \bibinfo{pages}{125801}
  (\bibinfo{year}{2009}).

\bibitem[{\citenamefont{Fujita et~al.}(2008)\citenamefont{Fujita, Jalil, and
  Tan}}]{fujita2008:JPCM}
\bibinfo{author}{\bibfnamefont{T.}~\bibnamefont{Fujita}},
  \bibinfo{author}{\bibfnamefont{M.~B.~A.} \bibnamefont{Jalil}},
  \bibnamefont{and} \bibinfo{author}{\bibfnamefont{S.~G.} \bibnamefont{Tan}},
  \bibinfo{journal}{J. Phys.: Condens. Matter} \textbf{\bibinfo{volume}{20}},
  \bibinfo{pages}{115206} (\bibinfo{year}{2008}).

\bibitem[{\citenamefont{Thomas et~al.}(1996)\citenamefont{Thomas, Nicholls,
  Simmons, Pepper, Mace, and Ritchie}}]{thomas1996:PRL}
\bibinfo{author}{\bibfnamefont{K.~J.} \bibnamefont{Thomas}},
  \bibinfo{author}{\bibfnamefont{J.~T.} \bibnamefont{Nicholls}},
  \bibinfo{author}{\bibfnamefont{M.~Y.} \bibnamefont{Simmons}},
  \bibinfo{author}{\bibfnamefont{M.}~\bibnamefont{Pepper}},
  \bibinfo{author}{\bibfnamefont{D.~R.} \bibnamefont{Mace}}, \bibnamefont{and}
  \bibinfo{author}{\bibfnamefont{D.~A.} \bibnamefont{Ritchie}},
  \bibinfo{journal}{Phys. Rev. Lett.} \textbf{\bibinfo{volume}{77}},
  \bibinfo{pages}{135} (\bibinfo{year}{1996}).

\bibitem[{\citenamefont{Chen et~al.}(2009)\citenamefont{Chen, Graham, Pepper,
  Sfigakis, Farrer, and Ritchie}}]{chen2009:PRB}
\bibinfo{author}{\bibfnamefont{T.-M.} \bibnamefont{Chen}},
  \bibinfo{author}{\bibfnamefont{A.~C.} \bibnamefont{Graham}},
  \bibinfo{author}{\bibfnamefont{M.}~\bibnamefont{Pepper}},
  \bibinfo{author}{\bibfnamefont{F.}~\bibnamefont{Sfigakis}},
  \bibinfo{author}{\bibfnamefont{I.}~\bibnamefont{Farrer}}, \bibnamefont{and}
  \bibinfo{author}{\bibfnamefont{D.~A.} \bibnamefont{Ritchie}},
  \bibinfo{journal}{Phys. Rev. B} \textbf{\bibinfo{volume}{79}},
  \bibinfo{pages}{081301(R)} (\bibinfo{year}{2009}).

\bibitem[{\citenamefont{Koop et~al.}(2007)\citenamefont{Koop, Lerescu, Liu,
  {van Wees}, Reuter, Wieck, and {van der Wal}}}]{koop2007:JSNM}
\bibinfo{author}{\bibfnamefont{E.}~\bibnamefont{Koop}},
  \bibinfo{author}{\bibfnamefont{A.}~\bibnamefont{Lerescu}},
  \bibinfo{author}{\bibfnamefont{J.}~\bibnamefont{Liu}},
  \bibinfo{author}{\bibfnamefont{B.}~\bibnamefont{{van Wees}}},
  \bibinfo{author}{\bibfnamefont{D.}~\bibnamefont{Reuter}},
  \bibinfo{author}{\bibfnamefont{A.}~\bibnamefont{Wieck}}, \bibnamefont{and}
  \bibinfo{author}{\bibfnamefont{C.}~\bibnamefont{{van der Wal}}},
  \bibinfo{journal}{J. Supercond. Nov. Magn.} \textbf{\bibinfo{volume}{20}},
  \bibinfo{pages}{433} (\bibinfo{year}{2007}).

\bibitem[{\citenamefont{Pallecchi et~al.}(2002)\citenamefont{Pallecchi, Heyn,
  Lohse, Kramer, and Hansen}}]{pallecchi2002:PRB}
\bibinfo{author}{\bibfnamefont{I.}~\bibnamefont{Pallecchi}},
  \bibinfo{author}{\bibfnamefont{C.}~\bibnamefont{Heyn}},
  \bibinfo{author}{\bibfnamefont{J.}~\bibnamefont{Lohse}},
  \bibinfo{author}{\bibfnamefont{B.}~\bibnamefont{Kramer}}, \bibnamefont{and}
  \bibinfo{author}{\bibfnamefont{W.}~\bibnamefont{Hansen}},
  \bibinfo{journal}{Phys. Rev. B} \textbf{\bibinfo{volume}{65}},
  \bibinfo{pages}{125303} (\bibinfo{year}{2002}).

\bibitem[{\citenamefont{MacLean et~al.}(2007)\citenamefont{MacLean, Amasha,
  Radu, {Zumb\"uhl}, Kastner, Hanson, and Gossard}}]{amasha2007:PRL}
\bibinfo{author}{\bibfnamefont{K.}~\bibnamefont{MacLean}},
  \bibinfo{author}{\bibfnamefont{S.}~\bibnamefont{Amasha}},
  \bibinfo{author}{\bibfnamefont{I.~P.} \bibnamefont{Radu}},
  \bibinfo{author}{\bibfnamefont{D.~M.} \bibnamefont{{Zumb\"uhl}}},
  \bibinfo{author}{\bibfnamefont{M.~A.} \bibnamefont{Kastner}},
  \bibinfo{author}{\bibfnamefont{M.~P.} \bibnamefont{Hanson}},
  \bibnamefont{and} \bibinfo{author}{\bibfnamefont{A.~C.}
  \bibnamefont{Gossard}}, \bibinfo{journal}{Phys. Rev. Lett.}
  \textbf{\bibinfo{volume}{98}}, \bibinfo{pages}{036802}
  (\bibinfo{year}{2007}).

\bibitem[{\citenamefont{Aleiner and Fa\v{l}ko}(2001)}]{aleiner2001:PRL}
\bibinfo{author}{\bibfnamefont{I.~L.} \bibnamefont{Aleiner}} \bibnamefont{and}
  \bibinfo{author}{\bibfnamefont{V.~I.} \bibnamefont{Fa\v{l}ko}},
  \bibinfo{journal}{Phys. Rev. Lett.} \textbf{\bibinfo{volume}{87}},
  \bibinfo{pages}{256801} (\bibinfo{year}{2001}).

\bibitem[{\citenamefont{Levitov and Rashba}(2003)}]{levitov2003:PRB}
\bibinfo{author}{\bibfnamefont{L.~S.} \bibnamefont{Levitov}} \bibnamefont{and}
  \bibinfo{author}{\bibfnamefont{E.~I.} \bibnamefont{Rashba}},
  \bibinfo{journal}{Phys. Rev. B} \textbf{\bibinfo{volume}{67}},
  \bibinfo{pages}{115324} (\bibinfo{year}{2003}).

\bibitem[{\citenamefont{Gurvitz and Kalbermann}(1987)}]{gurvitz1987:PRL}
\bibinfo{author}{\bibfnamefont{S.~A.} \bibnamefont{Gurvitz}} \bibnamefont{and}
  \bibinfo{author}{\bibfnamefont{G.}~\bibnamefont{Kalbermann}},
  \bibinfo{journal}{Phys. Rev. Lett.} \textbf{\bibinfo{volume}{59}},
  \bibinfo{pages}{262} (\bibinfo{year}{1987}).

\bibitem[{foo({\natexlab{a}})}]{footnote3}
\bibinfo{note}{For the structure geometry, our results depend only on the
  difference of the linear spin-orbit strengths $|l_{\rm br}^{-1}-l_{\rm
  d}^{-1}|$.}

\bibitem[{\citenamefont{Amasha et~al.}(2008{\natexlab{b}})\citenamefont{Amasha,
  MacLean, Radu, {Zumb\"uhl}, Kastner, Hanson, and Gossard}}]{amasha2008:PRL}
\bibinfo{author}{\bibfnamefont{S.}~\bibnamefont{Amasha}},
  \bibinfo{author}{\bibfnamefont{K.}~\bibnamefont{MacLean}},
  \bibinfo{author}{\bibfnamefont{I.~P.} \bibnamefont{Radu}},
  \bibinfo{author}{\bibfnamefont{D.~M.} \bibnamefont{{Zumb\"uhl}}},
  \bibinfo{author}{\bibfnamefont{M.~A.} \bibnamefont{Kastner}},
  \bibinfo{author}{\bibfnamefont{M.~P.} \bibnamefont{Hanson}},
  \bibnamefont{and} \bibinfo{author}{\bibfnamefont{A.~C.}
  \bibnamefont{Gossard}}, \bibinfo{journal}{Phys. Rev. Lett.}
  \textbf{\bibinfo{volume}{100}}, \bibinfo{pages}{046803}
  (\bibinfo{year}{2008}{\natexlab{b}}).

\bibitem[{\citenamefont{Hanson et~al.}(2003)\citenamefont{Hanson, Witkamp,
  Vandersypen, {van Beveren}, Elzerman, and Kouwenhoven}}]{hanson2003:PRL}
\bibinfo{author}{\bibfnamefont{R.}~\bibnamefont{Hanson}},
  \bibinfo{author}{\bibfnamefont{B.}~\bibnamefont{Witkamp}},
  \bibinfo{author}{\bibfnamefont{L.~M.~K.} \bibnamefont{Vandersypen}},
  \bibinfo{author}{\bibfnamefont{L.~H.~W.} \bibnamefont{{van Beveren}}},
  \bibinfo{author}{\bibfnamefont{J.~M.} \bibnamefont{Elzerman}},
  \bibnamefont{and} \bibinfo{author}{\bibfnamefont{L.~P.}
  \bibnamefont{Kouwenhoven}}, \bibinfo{journal}{Phys. Rev. Lett.}
  \textbf{\bibinfo{volume}{91}}, \bibinfo{pages}{196802}
  (\bibinfo{year}{2003}).

\bibitem[{\citenamefont{{Fa\v{l}ko} et~al.}(2005)\citenamefont{{Fa\v{l}ko},
  Altshuler, and Tsyplyatyev}}]{falko2005:PRL}
\bibinfo{author}{\bibfnamefont{V.~I.} \bibnamefont{{Fa\v{l}ko}}},
  \bibinfo{author}{\bibfnamefont{B.~L.} \bibnamefont{Altshuler}},
  \bibnamefont{and}
  \bibinfo{author}{\bibfnamefont{O.}~\bibnamefont{Tsyplyatyev}},
  \bibinfo{journal}{Phys. Rev. Lett.} \textbf{\bibinfo{volume}{95}},
  \bibinfo{pages}{076603} (\bibinfo{year}{2005}).

\bibitem[{\citenamefont{Kiselev et~al.}(1998)\citenamefont{Kiselev, Ivchenko,
  and {R\"ossler}}}]{kiselev1998:PRB}
\bibinfo{author}{\bibfnamefont{A.~A.} \bibnamefont{Kiselev}},
  \bibinfo{author}{\bibfnamefont{E.~L.} \bibnamefont{Ivchenko}},
  \bibnamefont{and}
  \bibinfo{author}{\bibfnamefont{U.}~\bibnamefont{{R\"ossler}}},
  \bibinfo{journal}{Phys. Rev. B} \textbf{\bibinfo{volume}{58}},
  \bibinfo{pages}{16353} (\bibinfo{year}{1998}).

\bibitem[{\citenamefont{Doty et~al.}(2006)\citenamefont{Doty, Scheibner,
  Ponomarev, Stinaff, Bracker, Korenev, Reinecke, and Gammon}}]{doty2006:PRL}
\bibinfo{author}{\bibfnamefont{M.~F.} \bibnamefont{Doty}},
  \bibinfo{author}{\bibfnamefont{M.}~\bibnamefont{Scheibner}},
  \bibinfo{author}{\bibfnamefont{I.~V.} \bibnamefont{Ponomarev}},
  \bibinfo{author}{\bibfnamefont{E.~A.} \bibnamefont{Stinaff}},
  \bibinfo{author}{\bibfnamefont{A.~S.} \bibnamefont{Bracker}},
  \bibinfo{author}{\bibfnamefont{V.~L.} \bibnamefont{Korenev}},
  \bibinfo{author}{\bibfnamefont{T.~L.} \bibnamefont{Reinecke}},
  \bibnamefont{and} \bibinfo{author}{\bibfnamefont{D.}~\bibnamefont{Gammon}},
  \bibinfo{journal}{Phys. Rev. Lett.} \textbf{\bibinfo{volume}{97}},
  \bibinfo{pages}{197202} (\bibinfo{year}{2006}).

\bibitem[{foo({\natexlab{b}})}]{footnote4}
\bibinfo{note}{Unpublished data from the experiments of Amasha {\it et al.}
  point against this first mechanism, S.~ Amasha and K.~ MacLean (private
  communication).}

\bibitem[{\citenamefont{Sablikov and Tkach}(2007)}]{sablikov2007:PRB}
\bibinfo{author}{\bibfnamefont{V.~A.} \bibnamefont{Sablikov}} \bibnamefont{and}
  \bibinfo{author}{\bibfnamefont{Y.~Y.} \bibnamefont{Tkach}},
  \bibinfo{journal}{Phys. Rev. B} \textbf{\bibinfo{volume}{76}},
  \bibinfo{pages}{245321} (\bibinfo{year}{2007}).

\bibitem[{\citenamefont{Lee and Bruder}(2005)}]{lee2005:PRB}
\bibinfo{author}{\bibfnamefont{M.}~\bibnamefont{Lee}} \bibnamefont{and}
  \bibinfo{author}{\bibfnamefont{C.}~\bibnamefont{Bruder}},
  \bibinfo{journal}{Phys. Rev. B} \textbf{\bibinfo{volume}{72}},
  \bibinfo{pages}{045353} (\bibinfo{year}{2005}).

\bibitem[{\citenamefont{Usaj and Balseiro}(2005)}]{usaj2005:EL}
\bibinfo{author}{\bibfnamefont{G.}~\bibnamefont{Usaj}} \bibnamefont{and}
  \bibinfo{author}{\bibfnamefont{C.~A.} \bibnamefont{Balseiro}},
  \bibinfo{journal}{Europhys. Lett.} \textbf{\bibinfo{volume}{72}},
  \bibinfo{pages}{631} (\bibinfo{year}{2005}).

\bibitem[{\citenamefont{Serra et~al.}(2007)\citenamefont{Serra, Sanchez, and
  Lopez}}]{serra2007:PRB}
\bibinfo{author}{\bibfnamefont{L.}~\bibnamefont{Serra}},
  \bibinfo{author}{\bibfnamefont{D.}~\bibnamefont{Sanchez}}, \bibnamefont{and}
  \bibinfo{author}{\bibfnamefont{R.}~\bibnamefont{Lopez}},
  \bibinfo{journal}{Phys. Rev. B} \textbf{\bibinfo{volume}{76}},
  \bibinfo{pages}{045339} (\bibinfo{year}{2007}).

\bibitem[{\citenamefont{Nguyen et~al.}(2009)\citenamefont{Nguyen, Drouhin,
  Wegrowe, and Fishman}}]{nguyen2009:PRB}
\bibinfo{author}{\bibfnamefont{T.~L.~H.} \bibnamefont{Nguyen}},
  \bibinfo{author}{\bibfnamefont{H.-J.} \bibnamefont{Drouhin}},
  \bibinfo{author}{\bibfnamefont{J.-E.} \bibnamefont{Wegrowe}},
  \bibnamefont{and} \bibinfo{author}{\bibfnamefont{G.}~\bibnamefont{Fishman}},
  \bibinfo{journal}{Phys. Rev. B} \textbf{\bibinfo{volume}{79}},
  \bibinfo{pages}{165204} (\bibinfo{year}{2009}).

\bibitem[{\citenamefont{Potok et~al.}(2003)\citenamefont{Potok, Folk, Marcus,
  Umansky, Hanson, and Gossard}}]{potok2003:PRL}
\bibinfo{author}{\bibfnamefont{R.~M.} \bibnamefont{Potok}},
  \bibinfo{author}{\bibfnamefont{J.~A.} \bibnamefont{Folk}},
  \bibinfo{author}{\bibfnamefont{C.~M.} \bibnamefont{Marcus}},
  \bibinfo{author}{\bibfnamefont{V.}~\bibnamefont{Umansky}},
  \bibinfo{author}{\bibfnamefont{M.}~\bibnamefont{Hanson}}, \bibnamefont{and}
  \bibinfo{author}{\bibfnamefont{A.~C.} \bibnamefont{Gossard}},
  \bibinfo{journal}{Phys. Rev. Lett.} \textbf{\bibinfo{volume}{91}},
  \bibinfo{pages}{016802} (\bibinfo{year}{2003}).

\bibitem[{\citenamefont{Meier et~al.}(2007)\citenamefont{Meier, Salis,
  Shorubalko, Gini, {Sch\"on}, and Ensslin}}]{meier2007:NP}
\bibinfo{author}{\bibfnamefont{L.}~\bibnamefont{Meier}},
  \bibinfo{author}{\bibfnamefont{G.}~\bibnamefont{Salis}},
  \bibinfo{author}{\bibfnamefont{I.}~\bibnamefont{Shorubalko}},
  \bibinfo{author}{\bibfnamefont{E.}~\bibnamefont{Gini}},
  \bibinfo{author}{\bibfnamefont{S.}~\bibnamefont{{Sch\"on}}},
  \bibnamefont{and} \bibinfo{author}{\bibfnamefont{K.}~\bibnamefont{Ensslin}},
  \bibinfo{journal}{Nat. Phys.} \textbf{\bibinfo{volume}{3}},
  \bibinfo{pages}{650} (\bibinfo{year}{2007}).

\bibitem[{\citenamefont{Peres}(1980)}]{peres1980:JPA}
\bibinfo{author}{\bibfnamefont{A.}~\bibnamefont{Peres}}, \bibinfo{journal}{J.
  Phys. A} \textbf{\bibinfo{volume}{13}}, \bibinfo{pages}{2979}
  (\bibinfo{year}{1980}).

\bibitem[{\citenamefont{Gurvitz}(1988)}]{gurvitz1988:PRA}
\bibinfo{author}{\bibfnamefont{S.~A.} \bibnamefont{Gurvitz}},
  \bibinfo{journal}{Phys. Rev. A} \textbf{\bibinfo{volume}{38}},
  \bibinfo{pages}{1747} (\bibinfo{year}{1988}).

\bibitem[{\citenamefont{Bardeen}(1961)}]{bardeen1961:PRL}
\bibinfo{author}{\bibfnamefont{J.}~\bibnamefont{Bardeen}},
  \bibinfo{journal}{Phys. Rev. Lett.} \textbf{\bibinfo{volume}{6}},
  \bibinfo{pages}{57} (\bibinfo{year}{1961}).

\bibitem[{\citenamefont{Wigner}(1955)}]{wigner1955:PR}
\bibinfo{author}{\bibfnamefont{E.~P.} \bibnamefont{Wigner}},
  \bibinfo{journal}{Phys. Rev.} \textbf{\bibinfo{volume}{98}},
  \bibinfo{pages}{145} (\bibinfo{year}{1955}).

\bibitem[{\citenamefont{Jalabert et~al.}(1992)\citenamefont{Jalabert, Stone,
  and Alhassid}}]{jalabert1992:PRL}
\bibinfo{author}{\bibfnamefont{R.~A.} \bibnamefont{Jalabert}},
  \bibinfo{author}{\bibfnamefont{A.~D.} \bibnamefont{Stone}}, \bibnamefont{and}
  \bibinfo{author}{\bibfnamefont{Y.}~\bibnamefont{Alhassid}},
  \bibinfo{journal}{Phys. Rev. Lett.} \textbf{\bibinfo{volume}{68}},
  \bibinfo{pages}{3468} (\bibinfo{year}{1992}).

\end{thebibliography}

\end{document}